\documentclass[pra,preprint,showpacs]{revtex4-1}
\usepackage[centertags]{amsmath}
\usepackage[koi8-r]{inputenc}
\usepackage{amsfonts}
\usepackage{amssymb}
\usepackage{amsthm} 
\usepackage{newlfont}
\usepackage{epsfig}
\usepackage{amscd}
\usepackage{graphicx}
\usepackage{epsfig}
\usepackage{footnote}
\usepackage{lipsum}
\usepackage{color}
\usepackage{xcolor}
\usepackage{graphicx}
\usepackage{multirow}
\usepackage{subcaption}
\usepackage{amscd}

\newcommand{\ZZ}{{\mathbb Z}}

\newcommand{\beq}{\begin{equation}}
\newcommand{\eeq}{\end{equation}}
\newcommand{\ba}{\begin{array}}
\newcommand{\ea}{\end{array}}
\newcommand{\bea}{\begin{eqnarray}}
\newcommand{\eea}{\end{eqnarray}}
\begin{document}
\begin{center}
{\large \sc \bf Two-level  control  {over quantum state creation}  via entangled equal-probability  state}

\vskip 15pt

S.I.Doronin, E.B.Fel'dman and A.I.Zenchuk

\vskip 8pt

{\it Federal Research Center of Problems of Chemical Physics and Medicinal Chemistry RAS,
Chernogolovka, Moscow reg., 142432, Russia}.

\end{center}

\begin{abstract}{
We propose the scheme realizing the two-level control over  the unitary operators $U_k$ creating the required quantum state of the system  $S$. These operators are controlled by the superposition state of the auxiliary subsystem $R$ which is governed by two  control centers.
The 
 first-level control center (main control)  creates the equal-probability pure state of $R$ with certain distribution of phase factors that, in turn,  govern the power of the second-level control center $C$ that applies the special $V$-operators  to the same subsystem $R$ changing its state and thus controlling the applicability of $U_k$. 
In addition,  the above phases are responsible for the entanglement in the subsystem $R$. We find the direct relation between this entanglement and   the  number of operators $U_k$ that can be controlled by $C$.  The simple example of a two-level control system governing the creation of entangled state of the two-qubit system  $S$ is presented.}
\end{abstract}


\maketitle

\section{Introduction}
\label{Section:Introduction}

\label{Section:state}

The control problems are of principal significance in the information theory. 
The control   {(in particular, remote control)} of quantum states can be  realized  by special control operations using, for instance, inhomogeneous interaction constants along the spin chain, controllable local magnetic field or controllable interaction with environment. 
For instance, the  NMR pulse sequence was used in \cite{KRKS} to maximize the coherence transfer and minimize the relaxation effects, 
quantum control involving both coherent (inhomogeneous magnetic field) or incoherent (interaction with environment) control was proposed in \cite{PP_2023}.
In both those cases the GRadient Ascent Pulse Engineering (GRAPE) was used to solve the optimization problem \cite{L,GV,FSGK}. 
Krotov method of optimization \cite{KF,K,MP_2019,FFLC}
was used in \cite{BM}
for optimizing quantum gates.
Quantum feedback control was used in \cite{SJL} for  information transfer
 and entanglement control \cite{MW}.
Quantum machine learning \cite{BWPRWL,NBSN,GSB}
represent another attractive field for implementing quantum control.
{
The controllable, robust, and fast entangled-state transmission scheme with the band-gap state of the non-Hermitian trimer Su-Schrieffer-Heeger  chain is proposed in \cite{CZCW}. The non-Hermiticity is induced by the periodic on-site imaginary potential on each cell.
Principles of non-Markovian quantum control are considered in \cite{NJS} and are  implemented to reinforcement learning involving the interaction with environment. 
Control of switching on a coupling term (undesired in many cases) between two quantum manybody
systems is considered in \cite{CBMF}.
The optimal control of a continuous variable system (in particular, entanglement control) using the Krotov
algorithm to optimize the state dynamics is considered in 
\cite{CLY}.
The multipartite entanglement properties in non-Hermitian superconducting qubits are studied in \cite{LFHJ}, where high-fidelity
entangled states can be created under strong driving fields or strong couplings among the qubits. 
The review of machine learning applications
to quantum communication
protocols (quantum key distribution, quantum teleportation, quantum secret sharing, and quantum networks) is given in \cite{M}.
}
 {Regarding remote state control (state transfer, state creation),} most profitable from the practical point of view are   photon systems providing the long distance communication 
\cite{PBGWK,PBGWK2,DLMRKBPVZBW}. Although  spin chains  are also of interest \cite{PSB,LH} and can be used in local communications between blocks of a particular device.

The main purpose of quantum control is creating the desired quantum states {satisfying certain requirements.} For that purpose it might be necessary  to organize the remote manipulation with 
controlling  operators. In fact, depending on the desired  state of the receiver  (in particular, on the dimensionality of the state) one can require application of different operators creating 
\cite{PBGWK,PBGWK2,DLMRKBPVZBW,PSB,LH,Z_2014,BZ_2015,SW,Licina,JWQCCLXSZM}
 or restoring \cite{FZ_2017,FPZ_2021,BFLPZ_2022,FPZ_2024} this state. 
{The two-level control proposed in this paper  assumes the hierarchy of control centers. The first (main) control center governs the power of the second control center. In particular, it can minimize this power, but cannot completely remove it. The second center establishes the control over  the state of considered system via switching on/off the appropriate  unitary transformations. }

{To clarify the principle of two-level control, we introduce several quantum subsystems.
The system $S$, whose state is of our interest, is subjected to the set of unitary transformations $U_k$ which are controlled by the superposition state of the auxiliary subsystem $R$. This subsystem is subjected to the action of both control centers.  
 The first (main) control center is represented by the one-qubit subsystem   $M$. Via the special unitary transformation, it creates  the equal probability superposition  state of 
 $R$ with phases distributed in a certain way.  Namely these phases define the ability of operators $V_i$ applied to $R$ and controlled by the qubits of $C$. Applying the operators $V_i$ allows the second-level center $C$ to switch off certain operators $U_k$ applied to $S$.
We show that the  control possibility of $C$ (number of operators $U_k$ that can be switched off) is directly related to the entanglement in the state of the auxiliary subsystem $R$  governed by $M$. More  exactly,
the subsystem $M$ generates  the entanglement  in the equal-probability state of $R$, and, in turn, this entanglement determines how many operators $U_k$ can be switched off by the $V$-operators controlled by $C$.}

The paper is organized as follows.
The protocol of the two-level control of a quantum state using  equal-probability state of the auxiliary subsystem $R$ and two control centers $M$ and $C$ is presented in Sec.\ref{Section:TwoLevel} together with appropriate circuit. { The particular example of two-level control protocol governing  creation of the entangled pure states of $S$ is presented.}  The entanglement in the bipartite  equal probability state and its relation to the efficiency of the control are studied in Sec.\ref{Section:Entanglement}. Basic conclusions are given in Sec.\ref{Section:Conclusions}. {The Appendix includes detailed  derivation of  the formula for  concurrence in the bipartite equal-probability state.}

\section{Two-level  control}
\label{Section:TwoLevel}
\begin{figure}[ht]
    \includegraphics[width=0.7\textwidth]{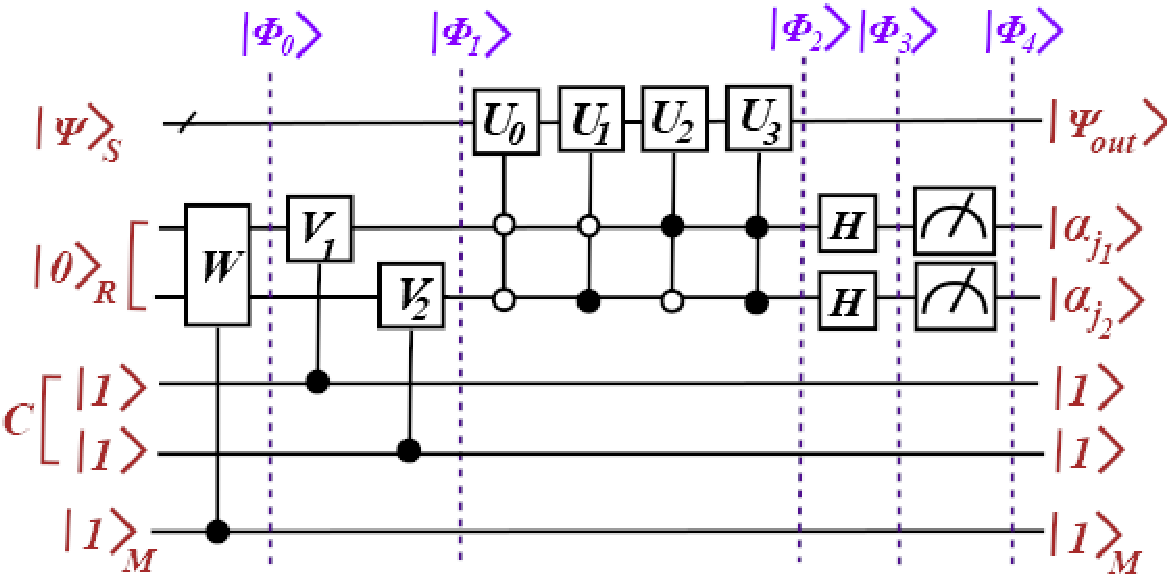}
    \caption{The application of equal-probability state (\ref{Psi}) in two-level control. In figure, $M$ is a one-qubit subsystem, $C$ and $R$ are both two-qubit subsystems.}
\label{Fig:SRCM}
\end{figure}

We consider the system $S$ subjected to action of the set of unitary transformations $U_k$. The final state of this system will be output of the protocol. We also  include three subsystems $R$, $C$ and $M$ serving to establish the two-level control over the operators $U_k$. This control is organized as follows, see  Fig.\ref{Fig:SRCM}.

{ The operators $U_k$  are controlled by the state of the subsystem $R$ which is $n^{(R)}$-qubit equal-probability pure state  $|\Psi_{eq.pr.}\rangle_R$:
\begin{eqnarray}\label{Psi}
|\Psi_{eq.pr.}\rangle_R =\frac{1}{{2^{n^{(R)}/2}}} \sum_{k=0}^{2^{n^{(R)}}-1} e^{i\varphi_k} |k\rangle,\;\;{\mbox{Im}}\varphi_k=0.
\end{eqnarray}
The general property of state (\ref{Psi}) is that measurement yields any state of the $n$-qubit system with the same probability $\frac{1}{2^{n}}$ for all $\varphi_k$. But entanglement in this state strongly depends on $\varphi_k$. 
In particular,  state (\ref{Psi}) with all $\varphi_i=0$ can be created applying the Hadamard transformation to each qubit of an $N$-qubit system in the ground state $|0\rangle$ and  therefore this state is not entangled.  Such state appears at the first step of the Phase Estimation Algorithm  \cite{NC}.  Another example of equal-probability state (\ref{Psi}) is the Quantum Fourier Transform:
\begin{eqnarray}
|j\rangle \to \frac{1}{\sqrt{2^n}} \sum_{k=0}^{2^n-1} e^{2\pi i j k/2^n} |k\rangle,
\end{eqnarray}
which is not entangled as well because it can be factorized into the one-qubit states \cite{NC}. 
}

{The phases in the equal probability state  (\ref{Psi}) are controlled by the operator $W$ applied to  the ground state  $|0\rangle_R$ of the subsystem  $R$ and controlled by $M$, see  Fig.\ref{Fig:SRCM}.} Thus, due to the equal probability state (\ref{Psi}) of $R$, all $U_k$ are applied to $S$ if only all $V_k$ are switched off. 
 In addition, the state of the subsystem $R$ is subjected to action of the operators $V_j$ controlled by the state of the subsystem $C$. These operators  can ''switch off '' some operators $U_k$ by changing the state of $R$. The set of operators $U_k$ that can be switched off by a given operator $V_j$ is defined not only by this operator but also by the phases in the equal-probability  state (\ref{Psi}) controlled by $W$. Thus, the subsystem $M$ establishes the indirect control over the action of the operators $V_j$, while $V_j$ represent the tool for the second-level control over the operators $U_k$  established by the system $C$. Note that after action of $V_j$ the state of $R$ loses its equal-probability property in general.

Thus, we deal with  the four-partite system $S-R-C-M$.
The relation between subsystems   is illustratively clarified in Fig.\ref{Fig:SRCM} for the two-qubit subsystems $R$, $C$ and 1-qubit subsystem $M$. 
All of $U_k$ are applied  to the subsystem $S$ except those that are switched off by the state of $R$ after applying operators  $V_k$. If all $V_k$ are identity operators, then all $U_k$ are applied to the subsystem $S$ since the state initiated by the operator $W$ is the equal-probability state.

{According to the above,} the number of operators $U_k$ that can be switched off by $V_k$ depends not only on the operator $V_k$ itself, but also on the phases $\varphi_k$ in the state (\ref{Psi}). We show that these phases define  the bipartite entanglement (concurrence) in the state of $R$ and the number of operators $U_k$ {controllable by} $V_k$ increases with decrease in the entanglement in $R$. This statement will be clarified below in more details. 

\subsection{Protocol for two-level control, Fig.\ref{Fig:SRCM}}
\label{Section:TLC}
Now we consider the  two-level control algorithm in more details assuming, for our convenience,  that the number of qubits in $R$ is $n^{(R)}$,
and the number of  operators $U_k$ is $N$ ($k=0,\dots,N-1$), $N=2^{n^{(R)}}$, see the circuit in Fig.\ref{Fig:SRCM},  { where $n^{(R)}=2$}. 

For simplicity and to focus on the basic features of protocol, we {consider that} the single qubit of the subsystem $M$ and all $n^{(C)}$ qubits of the subsystem $C$ {are} in the excited states, {i.e., the state of $C$ is $|2^{n^{(C)}}-1\rangle_C$.} {Let $W |0\rangle_R =|\Psi_{eq.pr.}\rangle_R$.} Then,
the action of the control-$W$ operator $W_{RM}=W  \otimes |1\rangle_M\,_M\langle 1| +  I_R
\otimes |0\rangle_M\,_M\langle 0| 
$ is as follows:
\begin{eqnarray}
{|\Phi_0\rangle} {= W_{RM} |\Psi_S\rangle|0\rangle_R  |2^{n^{(C)}}-1\rangle_C |1\rangle_M}= |\Psi_S\rangle |\Psi_{eq.pr.}\rangle_R  |2^{n^{(C)}}-1\rangle_C |1\rangle_M,
\end{eqnarray}
where $I_R$ is the identity operator acting on $R$, and $|\Psi_{eq.pr.}\rangle_R$ is defined in (\ref{Psi}). 

Assuming that each $V_k$ is controlled by the $k$th qubit $C^{(k)}$ of the subsystem $C$ we 
write the control-$V$ operators as $W_{RC^{(k)}}=V_k \otimes |1\rangle_{C^{(k)}}\, _{C^{(k)}}\langle 1 | 
+I_{C^{(k)}}\otimes  |0\rangle_{C^{(k)}}\, _{C^{(k)}}\langle 0 | 
$. We collect all these control operators in the operator
$W_{RC} =\prod_k  W_{RC^{(k)}}$. Applying $W_{RC}$ to ${|\Phi_0\rangle}$ we obtain:
\begin{eqnarray}
{|\Phi_1\rangle} = W_{RC}{ |\Phi_0\rangle} =
 |\Psi_S\rangle \prod_k V_k |\Psi_{eq.pr.}\rangle_R  |2^{n^{(C)}}-1\rangle_C |1\rangle_M . 
\end{eqnarray}
The state of $R$, after applying the $V$-operators, reads
\begin{eqnarray} \label{tPsi}
 \prod_k V_k |\Psi_{eq.pr.}\rangle_R = G_1^{-1}\sum_{j=0}^{2^{n^{(R)}}-1} a_j|j\rangle_R, \;\; G_1=\sqrt{\sum_j |a_j|^2},
\end{eqnarray}
 assuming that some of $a_j$ can be zero. 

Now we discuss the control-$U$ operators acting on $S$. Consider the control operator $W_j$, 
\begin{eqnarray}
W_j =   U_j \otimes |j\rangle_R  \, _R\langle j|+ I_S\otimes (I_R- |j\rangle_R  \, _R\langle j| ) ,\;\; j=0,\dots, 2^{n^{(R)}}-1.
\end{eqnarray}
It is obvious, that the product of all $W_j$ yields
\begin{eqnarray}
W^{(1)}_{{RS}} = \prod_{j=0}^{2^{n^{(R)}}-1} W_j  = 
\sum_{j=0}^{2^{n^{(R)}}-1}U_j\otimes |j\rangle_R \, _R\langle j|  .
\end{eqnarray}
Applying $W^{(1)}_{{RS}}$ to the state $|\Phi_1\rangle$ and using (\ref{tPsi}) we obtain
\begin{eqnarray}
|\Phi_2\rangle &=& W^{(1)}_{{RS}} {|\Phi_1 \rangle}{=W^{(1)}_{{RS}} G_1^{-1} \sum_{j=0}^{2^{n^{(R)}}-1} a_j |\Psi\rangle_S |j\rangle_R  |2^{n^{(C)}}-1\rangle_C |1\rangle_M}=\\\nonumber
&&
G_1^{-1}\sum_{j=0}^{2^{n^{(R)}}-1} a_jU_j |\Psi\rangle_S |j\rangle_R  |2^{n^{(C)}}-1\rangle_C |1\rangle_M .
\end{eqnarray}
Now we apply the Hadamard operator {$H$} to each qubit of the subsystem $R$ and denote the set of these  operators by $H_R$.
We have 
\begin{eqnarray}\label{Phi3}
|\Phi_3\rangle = H_R{ |\Phi_2 \rangle}=
G_1^{-1}\sum_{j=0}^{2^{n^{(R)}}-1} a_jU_j |\Psi\rangle_S H_R|j\rangle_R  |2^{n^{(C)}}-1\rangle_C |1\rangle_M .
\end{eqnarray}
Using the binary representation for $j=j_1\dots j_{n^{(R)}}$ in $|j\rangle_R$, we write
\begin{eqnarray}
&&
H_R|j_1\dots j_{n^{(R)}}\rangle_R =\frac{1}{2^{n^{(R)}/2}} (|0\rangle + (-1)^{j_1} |1\rangle) \dots   (|0\rangle + (-1)^{j_{n^{(R)}}} |1\rangle)=\\\nonumber
&&  {\frac{1}{2^{n^{(R)}/2}}  \sum_{\alpha_{j_1}=0}^1 \dots \sum_{\alpha_{j_{n^{(R)}}}=0}^1  (-1)^{ \alpha_{j_1} j_1 + \dots + \alpha_{j_{n^{(R)}}} j_{n^{(R)}}} | \alpha_{j_1}\dots \alpha_{j_{n^{(R)}}}\rangle_R }
\end{eqnarray}
If we measure states of all qubits of $R$ with the output $|\alpha_{j_i}\rangle$  (either $|0\rangle$ or $|1\rangle$) for the $i$th qubit, 
then we get
\begin{eqnarray}
&&
|\Phi_4\rangle ={|\Psi_{out}\rangle_{ \alpha_{j_1}\dots \alpha_{j_{n^{(R)}}}}} |2^{n^{(C)}}-1 \rangle_C |1\rangle_M ,\\\label{PsiOut}
&&
{|\Psi_{out}\rangle_{ \alpha_{j_1}\dots \alpha_{j_{n^{(R)}}}}} =\\\nonumber
&&G_2^{-1}
\sum_{j_1=0}^{1} \dots 
\sum_{j_{n^{(R)}}=0}^{1} \frac{1}{2^{n^{(R)}/2}}  (-1)^{ \alpha_{j_1} j_1 + \dots + \alpha_{j_{n^{(R)}}} j_{n^{(R)}}} a_{j_1\dots j_{n^{(R)}}}U_{j_1\dots j_{n^{(R)}}} |\Psi\rangle_S ,
\end{eqnarray}
where the normalization constant $G_2$ is defined by the  condition  {
$$_{ \alpha_{j_1}\dots \alpha_{j_{n^{(R)}}}}\langle \Psi_{out}|\Psi_{out}\rangle_{ \alpha_{j_1}\dots \alpha_{j_{n^{(R)}}}}=1.
$$
Note that $|\Psi_{out}\rangle_{ \alpha_{j_1}\dots \alpha_{j_{n^{(R)}}}}$ may be zero for some sets $ \alpha_{j_1},\dots, \alpha_{j_{n^{(R)}}}$. This means that probability to get qubits of $R$  in the appropriate sets of states $ \alpha_{j_1},\dots, \alpha_{j_{n^{(R)}}}$ after measurement is zero, i.e., there is not term proportional to  $|\alpha_{j_1}\dots \alpha_{j_{n^{(R)}}}\rangle_R$ in the state $|\Phi_3\rangle$ (\ref{Phi3}).}
It is interesting to note that the particular result of measurement over the subsystem $R$ effects only on  the phase factors ahead of the unitary transformations  $U_{j_1\dots j_{n^{(R)}}}$.
Under assumption that these phase factors are not important, we admit any of $2^{n^{(R)}}$ possible results of the measurement over the system $R$.
This is the case when we don't have to run the algorithm many times till  obtaining the desired result of measurement. Any result is acceptable.

\subsection{One- and multi-qubit operators $V_k$}
We study the mechanism of switching off the operators $U_k$ via applying the operators $V_k$ to $R$ and show that the number of switchable operators $U_k$ is directly related to the entanglement in $R$. 
First, we consider the simplest case of a one-qubit operator $V_1$ and then turn to the general case of a multiqubit operator.

\subsubsection{One-qubit operator $V_1$}
\label{Section:Vk1}
Since all qubits in $R$ are equivalent we consider the action of $V_1$ (applied to the first qubit of $R$  and controlled by the first qubit of $C$)  on the state $|\Psi_{eq.pr.}\rangle_R$ assuming that it is the one-qubit unitary operator   in the form
\begin{eqnarray}
V_1=
a_{00} |0\rangle \langle 0| + a_{01} |0\rangle \langle 1|+
a_{10} |1\rangle \langle 0| + a_{11} |1\rangle \langle 1|
\end{eqnarray}
with unitarity conditions
\begin{eqnarray}\label{unit}
&&
|a_{00}|^2+ |a_{01}|^2 =1,\;\;\;a_{00} a_{10}^*+a_{01} a_{11}^*=0,
 \;\;\; |a_{10}|^2+ |a_{11}|^2 =1.
\end{eqnarray}
We can rewrite $|\Psi_{eq.pr.}\rangle_R$ separating the states $|0\rangle$ and $|1\rangle$ of the first qubit:
\begin{eqnarray}\label{EPsiR}
|\Psi_{eq.pr.}\rangle_R=\frac{1}{2^{n^{(R)}/2}}  \left(|0\rangle \sum_{k=0}^{2^{n^{(R)}-1}-1} e^{i \varphi_{0k}}|k\rangle +
|1\rangle \sum_{k=0}^{2^{n^{(R)}-1}-1} e^{i \varphi_{1k}}|k\rangle\right) ,
\end{eqnarray}
where the subscript $k$ in $\varphi_{0k}$ and  $\varphi_{1k}$ enumerates states of the remaining $n^{(R)}-1$ qubits. Accordingly, we change the subscript in $U_k$: $U_k\to U_{ik}$, $i=0,1$.
The action of $V_1$ yields
\begin{eqnarray}\label{VPsi}
V_1|\Psi_{eq.pr.}\rangle_R &=&\frac{1}{2^{n^{(R)}/2}}  \left(
 |0\rangle \sum_{k=0}^{2^{n^{(R)}-1}-1}\Big(a_{00}  e^{i \varphi_{0k}}+
 a_{01}  e^{i \varphi_{1k}}\Big)|k\rangle +\right.\\\nonumber
&&
\left. |1\rangle  \sum_{k=0}^{2^{n^{(R)}-1}-1}\Big(a_{10}  e^{i \varphi_{0k}} +
 a_{11}  e^{i \varphi_{1k}}\Big)|k\rangle \right).
\end{eqnarray}
Notice that each term in  (\ref{EPsiR}) and, consequently, in   (\ref{VPsi}) handles 
the appropriate operator $U_{ik}$ acting on $S$. 
Thus, if we put some  term to zero, then we switch off  the action of the corresponding operator $U_{ik}$. 
Let, for some fixed $k$ (for instance, $k=0$),
\begin{eqnarray}\label{a00} &&
a_{00} e^{i\varphi_{00}} + a_{01}e^{i\varphi_{10}} =0\;\;\Rightarrow \;\;
a_{00}=-a_{01}e^{i(\varphi_{10}-\varphi_{00})},
\end{eqnarray}
which removes the term with $k=0$ from  the first part of  (\ref{VPsi}). 
The similar equality for another value $k=l$ (which would remove the $l$th term in the first part of  (\ref{VPsi}) in addition to the $0$th one) is possible if only 
\begin{eqnarray}\label{phi1}
&&\varepsilon_{01;0l} = \varphi_{10}-\varphi_{00} -( \varphi_{1l}-\varphi_{0l})=0,\;\;l\ge 1 \;\Rightarrow \\\label{phi2}
&&\varphi_{1k} =\varphi_{0k} +\varepsilon_{1},\;\;\;k=0,l,
\end{eqnarray} 
{where $\varepsilon_1$ is some parameter.}
Such $V_1$ cancels the terms with $k=0$ and $k=l$ {from the  first part of} (\ref{VPsi})  and thus switches off   $U_{00}$ and $U_{0l}$.
If the condition  (\ref{phi2}) holds for all $k$, then all operators $U_{0k}$, $k=0,\dots,2^{n^{(R)}-1}-1$,  (that are governed by   the first part of (\ref{VPsi})) will be switched off, and, moreover,   we can write (\ref{EPsiR}) as 
\begin{eqnarray}\label{R1}
|\Psi_{eq.pr.}\rangle_R=\frac{1}{2^{n^{(R)}/2}} (|0\rangle + e^{i\varepsilon_{1}}|1\rangle) \sum_{k=0}^{2^{n^{(R)}-1}-1} e^{i\varphi_{0k}} |k\rangle,
\end{eqnarray}
which means that the first qubit is not entangled with the rest of qubits in the state $|\Psi_{eq.pr.}\rangle_R$. This is the case when the operator $V_1$  {switches maximal number of} operators $U_{ik}$ {with $i=0$  and}  $k=1,\dots,2^{n^{(R)}}-1$.  Thus, we see that, controlling the phases in the probability amplitudes  of state  (\ref{EPsiR}), the subsystem $M$ controls the number of unitary operations $U_{0k}$ that can be switched off by $V_1$. Only if this number reaches the maximal value  (all terms in the first part of (\ref{VPsi}) can be set to 0) the  first qubit in state (\ref{EPsiR}) loses its entanglement with the rest of the  qubits of the subsystem $R$. Thus, adding constraints (\ref{phi1}) one by one 
we reduce this entanglement till it becomes  zero when the whole set of constraints (\ref{phi1}) is imposed. In general,  if we impose $N_j$ constraints  of type (\ref{phi1}), then the number of controlled operators $U_{0k}$ is $N_j+1$.

\paragraph{Explicit form of $V_1$.}
Now we define the elements  of the operator $V_1$ that allow to switch off the operators  $U_{0k}$. Substituting $a_{00}$ from (\ref{a00}) into the first and second equations in (\ref{unit}) we obtain 
\begin{eqnarray}\label{a11}
|a_{00}|^2=|a_{01}|^2=\frac{1}{2},\;\; a_{11} = a_{10} e^{i(\varphi_{00}-\varphi_{10})}.
\end{eqnarray}
Substituting $a_{11}$ from (\ref{a11})  and $a_{00}$ from (\ref{a00})  into the third equation  in (\ref{unit}) we  have
\begin{eqnarray}
|a_{11}|^2=|a_{10}|^2=\frac{1}{2}.
\end{eqnarray}
Thus, we take $a_{01} = - \frac{1}{\sqrt{2}} e^{i{\delta_1}}$, 
$a_{10} =  \frac{1}{\sqrt{2}} e^{i{\delta_2}}$, 
\begin{eqnarray}
V_1=\frac{1}{\sqrt{2}}\left(
\begin{array}{cc}
e^{i({\delta_1} +\varphi_{10} -\varphi_{00})} & - e^{i{\delta_1}}\cr
e^{i{\delta_2}}&e^{i({\delta_2} +\varphi_{00} -\varphi_{10})}
\end{array}
\right).
\end{eqnarray}
For  the control purpose we set ${\delta_j}=0$, $j=1,2$, so that
\begin{eqnarray}\label{V1}
V_1=\frac{1}{\sqrt{2}}\left(
\begin{array}{cc}
e^{i(\varphi_{10} -\varphi_{00})} & - 1\cr
1&e^{-i(\varphi_{10} -\varphi_{00})}
\end{array}
\right).
\end{eqnarray}

{Formally, adding more one-qubit operators $V_i$, $i=1,2,\dots$, can be considered as a particular case of  applying the multiqubit operator $V=\dots V_2 V_1$. Therefore we consider the multiqubit operator $V_1$ in the next subsection. }

\subsubsection{Multi-qubit operator $V_1$}
\label{Section:VkM}
We consider a particular $n_A$-qubit operator $V_1$ applied to the first $n_A$ qubits of $R$ and controlled by the first qubit of $C$. Thus, we 
split  the subsystem $R$ into the subsystems $A$ and $B$ having, respectively,  $n_A$ and $n_B$ qubits, $n_A+n_B=n^{(R)}$.
Therefore, we change the subscript in $U_k$: $U_k\to U_{l_Ak_B}$.
 It is also convenient to  
 enumerate phases in (\ref{Psi}) by the double index $k\to k_Ak_B$, where $0\le k_A \le 2^{n_A}-1$,  $0\le k_B \le 2^{n_B}-1$.
Then we can rewrite the state (\ref{Psi}) as
\begin{eqnarray}\label{Psi2}
|\Psi_{eq.pr.}\rangle_R = \frac{1}{\sqrt{N_AN_B}}\sum_{k_A=0}^{N_A-1} \sum_{k_B=0}^{N_B-1}
e^{i \varphi_{k_Ak_B}}|k_A\rangle |k_B\rangle, \;\;\;N_A=2^{n_A}, \;\;N_B=2^{n_B}.
\end{eqnarray}
We represent the $n_A$-qubit operator  $V_{1}$ as
\begin{eqnarray}\label{V2}
V_{1} = \sum_{l_A,k_A=0}^{N_A-1} a_{l_A k_A} |l_A\rangle\,\langle k_A|,\;\;{V_1 V_1^\dagger =I_A,}
\end{eqnarray}
{where $I_A$ is the identity operator, acting on the subsystem $A$.}
Applying this operator to (\ref{Psi2}) we obtain 
\begin{eqnarray}\label{VPsi2}
V_1 |\Psi_{eq.pr.}\rangle_R &=& \frac{1}{\sqrt{N_AN_B}}\sum_{l_A=0}^{N_A-1}\sum_{k_B=0}^{N_B-1}\left(\sum_{k_A=0}^{N_A-1}a_{l_A k_A}
e^{i \varphi_{k_Ak_B}}\right) |l_A\rangle   |k_B\rangle=\\\nonumber
&&\frac{1}{\sqrt{N_AN_B}}\sum_{l_A=0}^{N_A-1}\sum_{k_B=0}^{N_B-1}A_{l_Ak_B} |l_A\rangle   |k_B\rangle,
\end{eqnarray}
where 
\begin{eqnarray}
A_{l_Ak_B} = \sum_{k_A=0}^{N_A-1}a_{l_A k_A}
e^{i \varphi_{k_Ak_B}}.
\end{eqnarray}
For the fixed $l_A$,  $0\le l_A\le N_A-1$,
the system of $N_A-1$ equations,
\begin{eqnarray}\label{AB}
A_{l_Ak_B} =0,\;\; k_B=0,\dots, N_A-2,
\end{eqnarray}
is solvable   with respect to some $N_A-1$ variables out of $N_A$ variables $a_{l_Ak_A}$, $k_A=0,\dots,N_A-1$, for the proper choice of phases $\varphi_{k_Ak_B}$ such that
\begin{eqnarray}
{\mbox{rank}} (\Phi_0) = N_A-1,\;\;\;  \Phi_0= \{e^{i \varphi_{k_Ak_B}}: k_A =0,\dots, N_A-1, k_B =0,\dots, N_A-2\} .
\end{eqnarray}
Thus we can put zero up to $(N_A-1) $ terms in the transformed state $V_1|\Psi_{eq.pr.}\rangle_R$ and therefore we can switch off up to $(N_A-1)$ operators $U_{l_Ak_B}$ (acting on the system $S$
and governed by the states $|l_A\rangle|k_B\rangle$ ($k_B=0,\dots,N_B-2$, $l_A$ is fixed) of $R$)
 just via the single $l_A$th row of $V_1$.
We can put zero more terms if, adding one more equation to system (\ref{AB}), say, with $k_B=N_A-1$, we don't change the rank of  the $\Phi$-matrix, i.e.,
\begin{eqnarray}\label{PHI}
{\mbox{rank}} (\Phi_1) ={\mbox{rank}} (\Phi_0) = N_A-1,\;\;\;  \Phi_1= \{e^{i \varphi_{k_Ak_B}}: k_A,k_B =0,\dots, N_A-1\} {.}
\end{eqnarray}
This can be achieved if we arrange the linear dependence of at least two columns, for instance, $i_B$ and $j_B$:
\begin{eqnarray}\label{AB1}
&&
\varphi_{i_Ak_B} = \varphi_{j_Ak_B} +\varepsilon_{i_Aj_A} ,\;\; k_B=i_B, j_B,\;\; {  i_A,j_A = 0,\dots, N_A-1,\;\; \varepsilon_{i_Ai_A}=0}\Rightarrow\\\label{AB2}
&&
\varepsilon_{i_Aj_A;i_Bj_B} = \varphi_{i_Ai_B} - \varphi_{j_Ai_B} -( \varphi_{i_Aj_B} - \varphi_{j_Aj_B})=0.
\end{eqnarray}
Then we can set to zero  $ N_A$ terms in (\ref{VPsi2})  and therefore we can switch off   $N_A$ operators $U_{l_Ak_B}$, {$k_B=0,\dots,N_A-1$,} acting on $S$ and controlled by the states $|l_A\rangle
|k_B\rangle$ ($k_B=0,\dots, N_A-1$, $l_A$ is fixed) of $R$.  In general, if $M_B$ columns are linearly dependent, we can put zero up to $ (N_A+M_B-2)$ terms in (\ref{VPsi2}) and can switch off  the same number of operators $U_{l_Ak_B}$.
If $M_B=N_B$, then all columns in $\Phi$,
\begin{eqnarray}\label{PHI2}
 \Phi= \{e^{i \varphi_{k_Ak_B}}: k_A = 0,\dots,N_A-1,\;\;k_B =0,\dots, N_B-1\},
\end{eqnarray} 
 are linearly dependent. 
This is the case when  solving single equation  in system (\ref{AB}), {for instance,
\begin{eqnarray}\label{AlA0}
A_{l_A0}=0,
\end{eqnarray}}
  and thus providing  the zero probability amplitude for single state $|l_A\rangle|0_B\rangle$ { with fixed $l_A$}, we put to zero 
the  probability amplitudes for states $|l_A\rangle|k_B\rangle$, $k_B=1,\dots,N_B-1$, due to the linear dependence of columns in $\Phi$.
Moreover,  the state $|\Psi_{eq.pr.}\rangle_R$  in (\ref{Psi2}) can be factorized  in this  case.

In fact, in this case (\ref{AB1}) and (\ref{AB2}) hold for all $i_B$, $j_B$. 
Obviously, not all relations (\ref{AB2}) are independent. The linearly independent set corresponds to fixing $i_A=0$ and $i_B=0$:
\begin{eqnarray}\label{ABb} 
\varphi_{0k_B} &=& \varphi_{j_Ak_B} +\varepsilon_{0j_A} ,\;\; k_B=0,\,j_B \;\;\Rightarrow\\\label{ABb2}
\varepsilon_{0j_A;0j_B} &=& { \varphi_{j_Aj_B} -  \varphi_{0j_B}-(  \varphi_{j_A 0}- \varphi_{00} )}=0,\\\nonumber
&& 1\le j_A\le N_A-1,\;\; 1\le j_B\le N_B-1 .
\end{eqnarray}
Other $\varepsilon$'s can be expressed in terms of (\ref{ABb2}) as follows:
\begin{eqnarray}\label{lindep}
&&\varepsilon_{0j_A;i_Bj_B}= \varepsilon_{0j_A;0j_B}-\varepsilon_{0j_A;0i_B},\\\nonumber
&&\varepsilon_{i_Aj_A;0j_B}= \varepsilon_{0j_A;0j_B}-\varepsilon_{0i_A;0j_B},\\\nonumber
&&\varepsilon_{i_Aj_A;i_Bj_B}= \varepsilon_{0j_A;0j_B}-\varepsilon_{0j_A;0i_B}-\varepsilon_{0i_A;0j_B}+\varepsilon_{0i_A;0i_B}.
\end{eqnarray}
Thus, the independent set (\ref{ABb2})  includes 
\begin{eqnarray}
\label{K}
K=
(N_A-1) (N_B-1) 
\end{eqnarray}
constraints. 
We order constraints (\ref{ABb2}) as follows. 
\begin{eqnarray}\label{CzeroOrder}
&&E_1=\{\varepsilon_{0j_A;01}= 0 \mod 2 \pi,\;\; \; j_A=1,\dots, N_A-1\},\\\nonumber
&&E_2=\{\varepsilon_{0j_A;02}= 0 \mod 2 \pi,\;\; \; j_A=1,\dots, N_A-1\},\\\nonumber
&&\cdots\cdots\cdots\cdots\cdots\cdots\\\nonumber
&&
E_{N_B-1}=\{\varepsilon_{0j_A;0(N_B-1)}= 0 \mod 2 \pi,\;\; \; j_A=1,\dots, N_A-1\}. 
\end{eqnarray}
{Here we use  $\mod 2 \pi$ algebra because the phases $\varphi_k$ in (\ref{Psi}) and therefore the phases $\varphi_{i_A j_B}$, $\forall\, i_a,j_B$, are defined up to the term  $2\pi n$, $n\in \ZZ$.}
Each 
set $E_k$ provides linear dependence of the $k$th column on the $k_B=0$ column of $\Phi$ in (\ref{PHI2})  {and switches the operator  $U_{l_Ak}$, $l_A$ is fixed.}
If all $E_k$ are imposed, $k=1,\dots,N_B-1$, 
then  state (\ref{Psi2}) can be factorized as follows (compare with (\ref{R1})):
\begin{eqnarray}\label{Psi22}
|\Psi_{eq.pr.}\rangle_R = \frac{1}{\sqrt{N_AN_B}} \left(
|0\rangle + \sum_{k_A=1}^{N_A-1} e^{i \varepsilon_{0k_A}}|k_A\rangle\right)  \sum_{k_B=0}^{N_B-1}
e^{i \varphi_{0k_B}} |k_B\rangle.
\end{eqnarray}
Obviously, the subsystems $A$ and $B$ are  unentangled in this case.
It is interesting that conditions (\ref{ABb2})  gradually  reduce entanglement between subsystems $A$ and $B$ as will be demonstrated in Sec.\ref{Section:Ent2qub}. At that, if successive  constraint establishes a new linear dependency between columns, the entanglement jumps down deeply. We notice that, $N_A=2$ for the one-qubit subsystem $A$,   and each list $E_k$ includes single constraint which agrees with Sec.\ref{Section:Vk1}. 

Hereafter we consider the case when the $l_A$th row of $V_1$ solves single equation  in (\ref{AB}) (with $k_B=0$) and thus switches off single operators $U_{l_A 0}$. Then,  the main control over $U_{l_Ak_B}$ is  via the phases of the state of $R$. This is possible because the phases in the state  of $R$ control the number of linearly dependent columns  in the matrix  $\Phi$ (\ref{PHI2}).  
If each of  $L$ rows of $V_1$ solves one equation of form (\ref{AB}), then 
$V_1$ controls $L N_B$ operators with the maximum $(N_A-1) N_B$ operators ($L<N_A$, otherwise $\det(V_1)=0$).
 Recall that there are $N=N_AN_B$ operators $U_{l_Ak_B}$ in our protocol.  

{ 
Thus, single operator $V_1$ controls most of the operators $U_{i_A i_B}$. However, this control, based on the single constraint (\ref{AlA0}) on the elements of $V_1$ for  $l_A$th row, is not flexible enough. For instance, if for a particular $l_A=l_0$ Eq.(\ref{AlA0}) is not solved (i.e., $A_{l_A0}\neq 0)$, then we lose the control over  $N_B$ operators $U_{l_0 k_B}$, $k_B=0,\dots, ,N_B-1$.  In other words, we cannot establish the control over a single $U$-operator following this way. To make the control more flexible, we have to  allow more constraints on the elements of $V_1$, i.e., we have to replace Eq.(\ref{AlA0}) with more general Eq.(\ref{AB}). In addition, we have to remove some of constraints on the phases of $|\Psi_{eq.pr.}\rangle_R$. With the purpose of flexible control, we also can replace  single operator $V_1$ by  the set of independent operators controlled by different nodes of the subsystem $C$. In this case, each particular operator will be supplemented with appropriate constraints on the phases $\varphi_k$ and all constraints must be compatible. In addition, the constraints of type (\ref{AB}) on the elements of each subsequent operator will depend on the elements of all previous ones, in general. Such generalization can be performed in a straightforward way and is left beyond the scope of this paper.
}

{\subsection{Example: controlled creation of entangled states}
We turn to Fig.\ref{Fig:SRCM} and follow formulae in Sec.\ref{Section:TLC}.
We consider the two-qubit subsystems $S$, $R$ and $C$ and one-qubit subsystem $M$. In addition, we  set the ground initial states $|0\rangle_S$ and $|0\rangle_R $ for, respectively, $S$ and $R$ and excited initial states for the qubits of $C$ and for $M$, i.e., $|11\rangle_C$ and $|1\rangle_M$ respectively.
Let
\begin{eqnarray}
W|0\rangle_R = |\Psi_{eq.pr.}\rangle_R=\frac{1}{2}\left(e^{i \varphi_{00}}|00\rangle_R + e^{i \varphi_{01}}|01\rangle_R + e^{i \varphi_{10}}|10\rangle_R + 
e^{i \varphi_{11}}|11\rangle_R 
\right).
\end{eqnarray}
Then $|\Phi_0\rangle = |00\rangle_S |\Psi_{eq.pr.}\rangle_R|11\rangle_C|1\rangle_M$.

We  include single one-qubit  operator $V_1$ applied to the first qubit of $R$, this operator is   given in (\ref{V1}), but we replace $\varphi_{ij}\to\chi_{ij}$ to make it independent on phases $\varphi_{ij}$:
\begin{eqnarray}\label{V1Ex}
V_1=\frac{1}{\sqrt{2}}\left(
\begin{array}{cc}
e^{i(\chi_{10} -\chi_{00})} & - 1\cr
1&e^{-i(\chi_{10} -\chi_{00})}
\end{array}
\right).
\end{eqnarray}
Thus  $|\Phi_1\rangle = |00\rangle_S V_1 |\Psi_{eq.pr.}\rangle_R|11\rangle_C|1\rangle_M$.

Let $U_2=U_3=I_S$ and operators $U_{0} = \sigma^{(x)}_1$, $U_{1} = \sigma^{(x)}_2$ be applied to the first and second qubits of a two-qubit system $S$. 
The  operator $W^{(1)}_{RS}$ now reads:
\begin{eqnarray}
W^{(1)}_{RS}=U_0\otimes  |00\rangle_R \, _R\langle 00| + U_1 \otimes |01\rangle_R \,  _R\langle 01| + I_S \otimes |10\rangle_R \, _R\langle 10| + I_S \otimes |11\rangle_R  \,_R\langle 11|,
\end{eqnarray} 
where $I_S$ is the identity operator applied to $S$.
The pure states $|\Phi_2\rangle$, $|\Phi_3\rangle$ and $|\Phi_4\rangle$ can be calculated directly. The state $|\Psi_{out}\rangle_{ij}$ corresponds to the measurement of two qubits of $R$ in the states $|i\rangle$ and $|j\rangle$ respectively. 
We consider three cases.

{\it Case 1.} No special relations between the parameters $\chi_{ij}$ in $V_1$ and 
$\varphi_{ij}$ in $|\Psi_{eq.pr.}\rangle_R$.  In this case both operators $U_0$ and $U_1$ are applied to $S$ and we obtain the entangled state in the form 
$|\Psi_{out}\rangle_{ij}=\alpha_{ij}^{(1)} |00\rangle_S + \alpha_{ij}^{(2)} |01\rangle_S+ \alpha_{ij}^{(3)} |10\rangle_S$, $\sum_{k=1}^3 |\alpha_{ij}^{(k)}|^2 =1$ with $\alpha_{ij}^{(k)}$, $k=1, 2, 3$, depending on $\varphi_{ij}$ and $\chi_{ij}$. In particular, if
$\varphi_{01} = \varphi_{00}$, $\varphi_{11}=\varphi_{10}$, 
$\varphi_{10} = \pi + \varphi_{00} +\chi_{10} - \chi_{00}$ we get the maximally entangled Bell states:
\begin{eqnarray}
&&
|\Psi_{out}\rangle_{00} =|\Psi_{out}\rangle_{10} = \alpha   (  |10\rangle_S+|01\rangle_S) ,\\\nonumber
&&
|\Psi_{out}\rangle_{01} =|\Psi_{out}\rangle_{11} =\alpha   (|10\rangle_S- |01\rangle_S) ,\\\nonumber
\end{eqnarray}
where
$\alpha= \frac{e^{i(\varphi_{00} -\chi_{00}+\chi_{10})}}{\sqrt{2}}$. We emphasize that the maximally entangled state appears at any result of measuring the state of $R$.

  {\it Case 2.}
  Condition (\ref{a00}) is satisfied, i.e., 
   $\chi_{00}=\varphi_{00}$, $\chi_{10}=\varphi_{10}$, only the operator $U_1$ is applied to $S$. We have the unentangled state in the form
   $|\Psi_{out}\rangle_{ij}=\alpha_{ij}^{(1)}  |00\rangle_S + \alpha_{ij}^{(2)} |01\rangle_S$, $|\alpha_{ij}^{(1)}|^2 + |\alpha_{ij}^{(2)}|^2=1$, 
   with $\alpha_{ij}^{(k)}$, $k=1,2$, depending on $\varphi_{ij}$. 
   In particular, if $\varphi_{01} =\pi + \varphi_{00}$, $\varphi_{11}=\pi + \varphi_{10}$, we obtain the ground state:
   \begin{eqnarray}
   &&
   |\Psi_{out}\rangle_{00} = |\Psi_{out}\rangle_{10}=0,\\\nonumber
&&
    |\Psi_{out}\rangle_{01}=-|\Psi_{out}\rangle_{11} = e^{i \varphi_{00}} |00\rangle_S.
   \end{eqnarray}
Again, the ground state appears  at any result of measuring the state of $R$.

{\it Case 3.} Condition (\ref{a00}) and constraint (\ref{phi1}) (with $l=1$) are satisfied, i.e.,
$\chi_{00}=\varphi_{00}$, $\chi_{10}=\varphi_{10}$, 
$\varphi_{11} = \varphi_{10} -\varphi_{00} + \varphi_{10}$. 
In this case no $U$-operators are applied to $S$. We obtain
\begin{eqnarray}
&&
|\Psi_{out}\rangle_{00} =- |\Psi_{out}\rangle_{10} = \frac{ e^{i\varphi_{00}} + e^{i\varphi_{01}}}{2} |00\rangle_S,\\\nonumber
&&
|\Psi_{out}\rangle_{01} =-|\Psi_{out}\rangle_{11} = \frac{ e^{i\varphi_{00}} - e^{i\varphi_{01}}}{2} |00\rangle_S,
\end{eqnarray}
i.e., we have the initial ground state up to the phase, similar to  Case 2.
   }

\section{Entanglement in the equal-probability state}
\label{Section:Entanglement}

{
We see that applying the $n_A$-qubit operator $V_1$ splits  $R$ into two subsystems $A$ and $B$ of $n_A$ and $n_B=n^{(R)}-n_A$ qubits respectively.  This fact prompts us to consider the entanglement in the bipartite system $A\cup B$.
}

The entanglement in the pure state $|\Psi\rangle$ of a bipartite system $A\cup B$ can be calculated by the formula \cite{BBPS}
\begin{eqnarray}
E=-{\mbox{Tr}}(\rho\log_2 \rho),
\end{eqnarray}
where $\rho$ is the density matrix obtained by  the partial trace of the density matrix $|\Psi\rangle\langle\Psi|$ over one of subsystems, either $A$ or $B$.
{However, in this paper, we use concurrence as a measure of entanglement  in a pure state of a bipartite system applying the following generalization of the Wootters criterion   \cite{HW,Wootters} proposed in \cite{RBCHM}:}
\begin{eqnarray}\label{C}
C(|\Psi\rangle) = \sqrt{2 (1-{\mbox{Tr}}(\rho_A)^2)},\;\; \rho_A = {\mbox{Tr}}_B |\Psi\rangle \langle \Psi|,
\end{eqnarray}
where ${\mbox{Tr}}_B$ means  partial trace  over the subsystem $B$.  This formula can be given a simple form for the case of an equal probability state which is Eq.(\ref{Psi2}) for a bi-partite system. 
{In this case
formula (\ref{C}) yields
\begin{eqnarray}\label{Cvar}
C=\frac{4}{N_AN_B}\sqrt{\sum_{i_A=0}^{N_A-2} 
  \sum_{i_B=0}^{N_B-2} 
 \sum_{j_A=i_A+1}^{N_A-1}  
 \sum_{j_B=i_B+1}^{N_B-1}  
  \sin^2\frac{\varepsilon_{i_Aj_A;i_Bj_B}} {2}},
\end{eqnarray}
where $\varepsilon_{i_Aj_A;i_Bj_B}$ is defined in (\ref{AB2}), $\varepsilon_{00;i_Bj_B}=\varepsilon_{i_Aj_A;00}=0$.  Details of derivation of Eq. (\ref{Cvar}) are given in Appendix, Sec. \ref{Section:Appendix}.
}

The concurrence is zero if each term under the square root  is zero, i.e.
\begin{eqnarray}\label{Czero0}
&&\varepsilon_{i_Aj_A;i_Bj_B}
 = 0 \mod 2 \pi,\\\nonumber
&&  i_A=0,\dots,N_A-2, \;\; j_A=i_A+1,\dots, N_A-1 ,\;\;\\\nonumber
&&
i_B =0,\dots, N_B-2,\;\;
j_B=i_B+1,\dots, N_B-1.
\end{eqnarray}
Obviously, not all conditions (\ref{Czero0}) are independent due to the relations (\ref{lindep})  among $\varepsilon_{i_Aj_A;i_Bj_B}$.
The set of independent conditions in list (\ref{Czero0})  is given  in (\ref{ABb2}).
It is remarkable that constraints (\ref{Czero0}) coincide with constraints (\ref{AB2}) providing linear dependency of the columns of the matrix $\Phi$ (\ref{PHI2}).
We order these constraints as in (\ref{CzeroOrder}).
Each set $E_k$ provides linear dependence of the $k$th column on the $k_B=0$ column of $\Phi$ (\ref{PHI2}).
Therefore, adding successive set of constraints $E_k$ we significantly reduce the entanglement between the subsystems $A$ and $B$, which is demonstrated  in Figs.\ref{Fig:NA2}, \ref{Fig:NA3} for examples of two- and three-qubit subsystem $A$, $n_A=2,3$.

Below we present detailed study of { concurrence (\ref{Cvar}) for the case of  one, two and three-qubit subsystem $A$} ($n_A=1,2,3$). 
We concentrate on the dependence of the concurrence on the number $N_K$ of constraints (\ref{CzeroOrder}) imposed on the phases $\varphi_{i_Aj_B}$ and, for a fixed $n_A$, introduce the function $C_{max}(N_K)$ by the formula
  \begin{eqnarray}
  \label{CmaxNK}
C_{max}(N_K) = \max_\varepsilon\, C(N_K,{\varepsilon}),
  \end{eqnarray}
where maximization is over those parameters $\varepsilon$, {
\begin{eqnarray}\label{ve}
\varepsilon = \{ \varepsilon_{0i_A;0j_B}: 1\le i_A\le N_A-1,\; 1\le i_B\le N_B-1\}
\end{eqnarray}}
that remain free after imposing   $N_K$ constraints out of the list (\ref{CzeroOrder}). {In (\ref{CmaxNK}),  we introduce two variables $N_K$ and $\varepsilon$ for the concurrence $C$ defined in (\ref{Cvar}).} 
{ The number of constraints $N_K$  is calculated as follows.  First, we add the constraints from the list $E_1$, then the constraints from the list $E_2$ and so on. Thus, $N_K$ constraints mean that we  take $p=\left[\frac{N_K}{N_A-1}\right]$ lists $E_i$ ($i=1,\dots p$) and $j=N_K-p$ first constraints from the list $E_{p+1}$.}

\subsection{One-qubit subsystem $A$}
\label{Section:NA1}
\begin{figure}[ht]
\centering
  \begin{subfigure}[c]{0.45\textwidth}
    \centering
    \includegraphics[width=\textwidth]{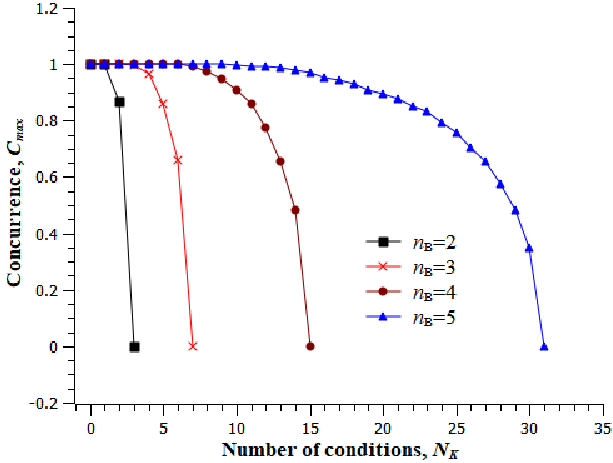}
    \caption{}
  \end{subfigure}
\hfill
\centering
  \begin{subfigure}[c]{0.45\textwidth}
    \centering
    \includegraphics[width=\textwidth]{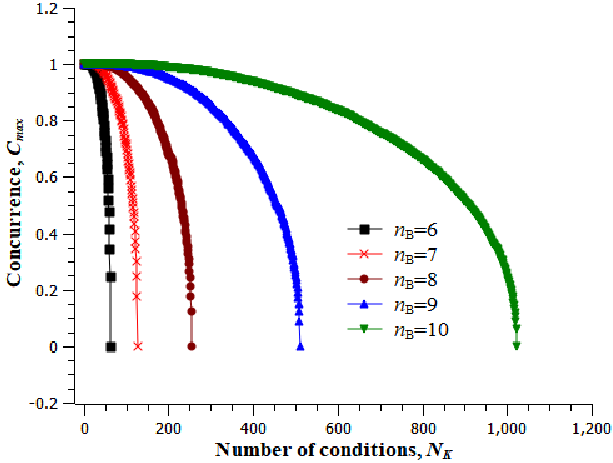}
    \caption{}
  \end{subfigure}
\caption{The dependence of  $C_{max}$ on the number $N_K$ of constraints (\ref{CzeroOrder1})  for different numbers of qubits $n_B$  in the subsystem $B$: {(a) $n_B=2$, 3, 4, 5; (b) $n_B=6$, 7, 8, 9, 10}. The case  $n_B=1$ consists of two points $C_{max}=1$ ($N_K=0$) and $C_{max}=0$ ($N_K=1$) and therefore it is not shown in this Figure. }
\label{Fig:NA1}
\end{figure}

In the case of the one-qubit subsystem $A$ {we have $n^{(A)}=1$, $n^{(B)}=N^{(R)}-1$,} formula (\ref{Cvar}) becomes simpler ($N_A=2$, $N_B=2^{n^{(R)}-1}$):
\begin{eqnarray}\label{Cvar1}
C=\frac{2}{N_B}\sqrt{
  \sum_{i_B=0}^{N_B-2}   
 \sum_{j_B=i_B+1}^{N_B-1}  
  \sin^2\frac{\varepsilon_{01;i_Bj_B}}{2}},
\end{eqnarray}
The maximum of $C$ equals 1 (this is confirmed in Fig.\ref{Fig:NA1})  and corresponds to 
\begin{eqnarray}
\varepsilon_{01;i_Bj_B}=\frac{2 \pi (i_B-j_B)}{N_B}.
\end{eqnarray}
The parameter $K=N_B-1$ in (\ref{K}).
The set of  ordered constraints  (\ref{CzeroOrder}) 
now reads
\begin{eqnarray}\label{CzeroOrder1}
&&E_1=\{\varepsilon_{01;01}= 0 \mod 2 \pi\},\\\nonumber
&&E_2=\{\varepsilon_{01;02}= 0 \mod 2 \pi\},\\\nonumber
&&\cdots\cdots\cdots\cdots\cdots\cdots\\\nonumber
&&
E_{N_B-1}=\{\varepsilon_{01;0(N_B-1)}= 0 \mod 2 \pi\}. 
\end{eqnarray}
We rewrite (\ref{Cvar1}) selecting the  terms associated with  constraints (\ref{CzeroOrder1}):
\begin{eqnarray}\label{Cvar2}
C=\frac{2}{N_B}\sqrt{ 
 \sum_{j_B=1}^{N_B-1}  
  \sin^2\frac{\varepsilon_{01;0j_B}} {2}
+
  \sum_{i_B=1}^{N_B-2}   
 \sum_{j_B=i_B+1}^{N_B-1}  
  \sin^2\frac{\varepsilon_{01;i_Bj_B}} {2}
},
\end{eqnarray}
and  grade the concurrence by the number $N_K$ of conditions (\ref{CzeroOrder})  imposed on  phases. 

1. One constraint: $N_K=1$, 
\begin{eqnarray}\label{constr1}
E_1: \;\varepsilon_{01;01} = 0 .
\end{eqnarray}
The first of Eqs.(\ref{lindep}) yields $\varepsilon_{01;11} =0$, and $\varepsilon_{01;1j_B} = \varepsilon_{01;0j_B}$, $j_B=2,\dots, N_B-1$. Then
\begin{eqnarray}\label{Cvar21}
C{(1,\varepsilon)}=\frac{2}{N_B}\sqrt{
2 \sum_{j_B=2}^{N_B-1}  
  \sin^2\frac{\varepsilon_{01;0j_B}} {2}
+
 \sum_{i_B=2}^{N_B-2}   
 \sum_{j_B=i_B+1}^{N_B-1}  
  \sin^2\frac{\varepsilon_{01;i_Bj_B}} {2}
},
\end{eqnarray}
{ where $\varepsilon$ is the list in (\ref{ve}).}

2. Two constraints: $N_K=2$, 
\begin{eqnarray}\label{constr2}
E_1, E_2:\; \varepsilon_{01;0j_B} = 0 ,\;\; j_B=1,2.
\end{eqnarray}
The first of Eqs.(\ref{lindep}) yields $\varepsilon_{01;1j_B} =0$, $j_B=1,2$, and $\varepsilon_{01;1j_B} = { \varepsilon_{01;2j_B} } =  \varepsilon_{01;0j_B}$,  $j_B=3,\dots, N_B-1$. Then
\begin{eqnarray}\label{Cvar3}
C{(2,\varepsilon)}=\frac{2}{N_B}\sqrt{
3 \sum_{j_B=3}^{N_B-1}  
  \sin^2\frac{\varepsilon_{01;0j_B}} {2}
+
 \sum_{i_B=3}^{N_B-2}   
 \sum_{j_B=i_B+1}^{N_B-1}  
  \sin^2\frac{\varepsilon_{01;i_Bj_B}} {2}
}.
\end{eqnarray}

3. $k$ constraints: $N_K=k$, 
\begin{eqnarray}\label{constr3}
E_1,\dots,E_k:\;\varepsilon_{01;0j_B}= 0 , \;\;j_B=1,\dots,k,\;\; k\le N_B-1.
\end{eqnarray}
The first of Eqs.(\ref{lindep}) yields $\varepsilon_{01;1j_B} =0$,   $j_B=1,\dots,k$, and  { $\varepsilon_{01;i_Bj_B} = \varepsilon_{01;0j_B}$,  $i_B=1,\dots,k$,} $j_B=k+1,\dots, N_B-1$. Then
\begin{eqnarray}\label{Cvar4}
C{(k,\varepsilon)}=\frac{2}{N_B}\sqrt{
(k+1) \sum_{j_B=k+1}^{N_B-1}  
  \sin^2\frac{\varepsilon_{01;0j_B}} {2}
+
 \sum_{i_B=k+1}^{N_B-2}   
 \sum_{j_B=i_B+1}^{N_B-1}  
  \sin^2\frac{\varepsilon_{01;i_Bj_B}} {2}
}.
\end{eqnarray}

If $k=N_B-2$, then $C{(N_B-2,\varepsilon)}=\frac{2}{N_B}\sqrt{(N_B-1)\sin^2\frac{\varepsilon_{01;0(N_B-1)}} {2} }$, and if $k=N_B-1$, then $C{(N_B-1,\varepsilon)}=0$.

Dependence of maximal  concurrence $C_{max}$, defined in (\ref{CmaxNK}), on the number $N_K$ of constraints (\ref{CzeroOrder1})
for  different numbers of  qubits in the subsystem $B$ is demonstrated in Fig.\ref{Fig:NA1}. 
The maximization is performed over 1000 random choices  (over 500 choices for $n_B=10$)  of the parameters  $\varepsilon_{01;0j_B}$, $j_B=N_K+1,\dots, N_B-1$  (which remain free after imposing $N_K$ constraints (\ref{CzeroOrder1})), $0\le \varepsilon_{01;0j_B} < 2 \pi$. In this case each constraint (\ref{CzeroOrder1})
introduces  a new column linearly dependent on the $0$th column in the matrix $\Phi$ (\ref{PHI2}). Therefore, for each $n_B$,  all values of concurrence  fall on the smooth curve. Another feature of this case is that $C_{max}=1$ at $N_K=0$ and does not depend on $n_B$. We notice also, that the maximal number of operators $U_{l_Ak_B}$ controllable by the subsystem $C$ is $N_B$, {where $l_A$ is 0 (or 1), in accordance with the arguments in the end of Sec.\ref{Section:VkM}.}

\subsection{Two qubit subsystem $A$}
\label{Section:Ent2qub}
\begin{figure}[ht]
\centering
  \begin{subfigure}[c]{0.45\textwidth}
    \centering
    \includegraphics[width=\textwidth]{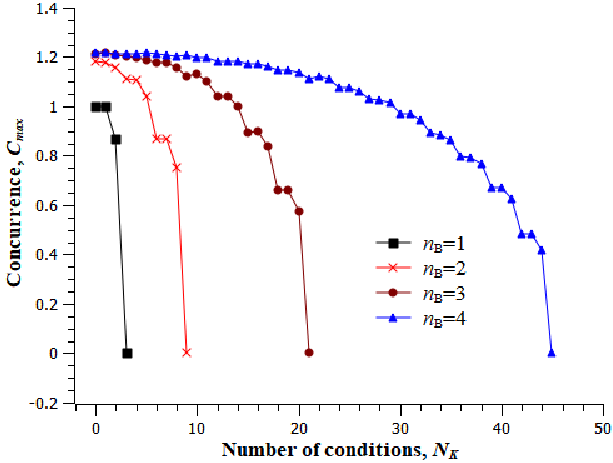}
    \caption{}
  \end{subfigure}
\hfill
\centering
  \begin{subfigure}[c]{0.45\textwidth}
    \centering
    \includegraphics[width=\textwidth]{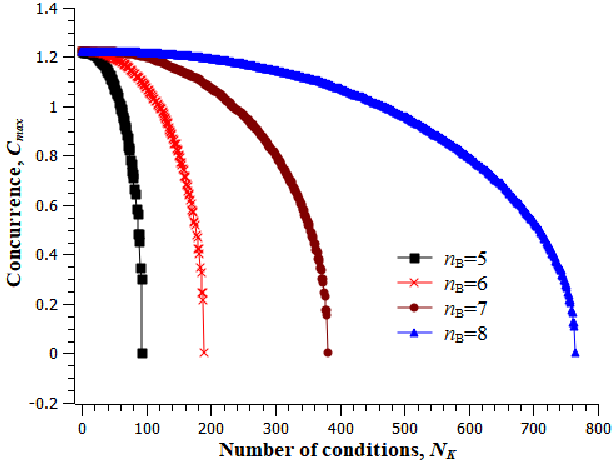}
    \caption{}
  \end{subfigure}
\caption{$n_A=2$; the dependence of $C_{max}$ on the number $N_K$ of constraints (\ref{CzeroOrder2})  for different numbers of qubits $n_B$  in the subsystem $B$: {(a) $n_B=1$, 2, 3, 4; (b) $n_B=5$, 6, 7, 8}.
}
\label{Fig:NA2}
\end{figure}

Similarly, we can consider the bi-partite concurrence in the case of two-qubit subsystem $A$: { $n_A=2$, $n_B=n^{(R)}-2$}.
The parameter $K=3  (N_B-1)$, $N_A=4$, {$N_B=2^{n^{(R)}-2}$}.
We study the dependence of concurrence on the number of constraints {$N_K$} imposed on the phases. The set of 
 ordered  constraints (\ref{CzeroOrder}) now reads:
\begin{eqnarray}\label{CzeroOrder2}
&&E_1=\{\varepsilon_{0j_A;01}
 = 0, \;\;j_A=1,2,3\}, \\\nonumber
&&E_2= \{ \varepsilon_{0j_A;02}
 = 0, \;\; j_A=1,2,3\}, \\\nonumber
&& \dots \dots   \dots \dots   \dots \dots    \\\nonumber
&& E_{N_B-1} = \{\varepsilon_{0j_A;0N_B-1}
 = 0, \;\; j_A=1,2,3\}.
\end{eqnarray}
Then (\ref{Cvar}) reads:
\begin{eqnarray}\label{Cvar00}
&&C=\frac{1}{N_B}\sqrt{\sum_{i_A=0}^{2} 
  \sum_{i_B=0}^{N_B-2} 
 \sum_{j_A=i_A+1}^{3}  
 \sum_{j_B=i_B+1}^{N_B-1}  
  \sin^2\frac{\varepsilon_{i_Aj_A;i_Bj_B}} {2}}{.}
  \end{eqnarray}
Dependence of the maximal  concurrence $C_{max}$, defined in (\ref{CmaxNK}), on the number  $N_K$ of constraints (\ref{CzeroOrder2})
for  different numbers $N_B$ of  qubits in the subsystem $B$ is demonstrated in Fig.\ref{Fig:NA2}.
Similar to Sec.\ref{Section:NA1}, the maximization is performed over 1000 random choices of those parameters   $\varepsilon_{0j_A;0j_B}$ that  remain free after imposing $N_K$ constraints  from set (\ref{CzeroOrder2}), $0\le \varepsilon_{0j_A;0j_B} < 2 \pi$.   Unlike the case $n_A=1$ in Sec. \ref{Section:NA1}, the curves are not smooth, the
 deeper jumps  down are observed in passing from 
{$N_{3k-1}$ constraints to $N_{3k}$ constraints, $k=1,\dots,N_B-1$}, since imposing  the set of three constraints  $E_k$  we establish linear dependence of the $k$th column of the matrix $\Phi$ (\ref{PHI2}) on the $0$th column.  {Such jumps are evident in Fig.\ref{Fig:NA2}a and are distinguishable in Fig.\ref{Fig:NA2}b for large $N_K$ (the lower parts of the curves).} In this case, the maximal number of operators $U_{l_Ak_B}$ controllable by the subsystem $C$ is $3 N_B$, {in accordance with the arguments given in the end of Sec.\ref{Section:VkM}}. Fig. \ref{Fig:NA2} shows that,
unlike Sec.\ref{Section:NA1}, $C_{max}$ at $N_K=0$ depends on $n_B$, and this dependence is demonstrated by the lower solid line  in Fig.\ref{Fig:max}. 

\subsection{Three qubit subsystem $A$}
\label{Section:Ent3qub}
\begin{figure}[ht]
\centering
  \begin{subfigure}[c]{0.45\textwidth}
    \centering
    \includegraphics[width=\textwidth]{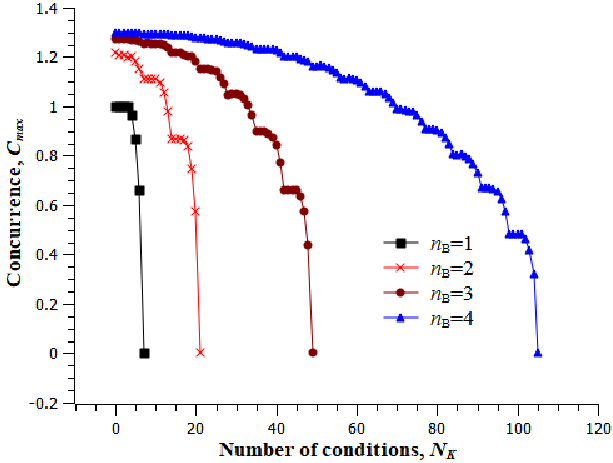}
    \caption{}
  \end{subfigure}
\hfill
\centering
  \begin{subfigure}[c]{0.45\textwidth}
    \centering
    \includegraphics[width=\textwidth]{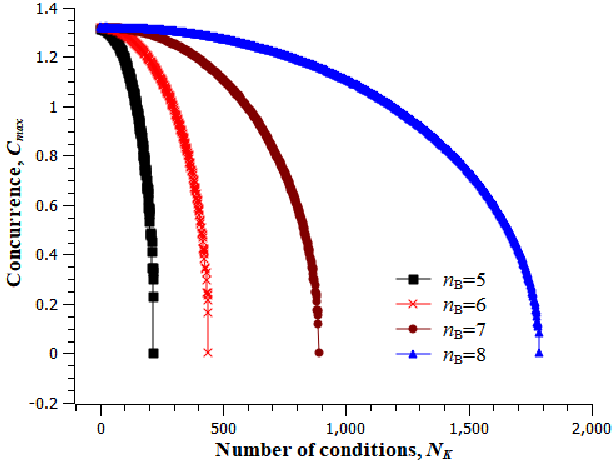}
    \caption{}
  \end{subfigure}
\caption{$n_A=3$; the dependence of $C_{max}$ on the number $N_K$ of constraints (\ref{CzeroOrder3})  for different numbers of qubits $n_B$ in the subsystem $B$: {(a) $n_B=1$, 2, 3, 4; (b) $n_B=5$, 6, 7, 8}. }
\label{Fig:NA3}
\end{figure}

Finally, we consider the bipartite concurrence in the case of three-qubit subsystem $A$. Now $n^{(A)}=3$, $n_B=n^{(R)}-3$, 
the parameter $K=7  (N_B-1)$, $N_A=8$, {$N_B=2^{n^{(R)}-3}$}.
Similar to Secs.\ref{Section:NA1} and \ref{Section:Ent2qub},  we study the dependence of concurrence on the number $N_K$ of constraints, imposed on the phases. The set of 
 ordered  constraints  (\ref{CzeroOrder}) now reads:
\begin{eqnarray}\label{CzeroOrder3}
&&E_1=\{\varepsilon_{0j_A;01}
 = 0, \;\;j_A=1,\dots,7\}, \\\nonumber
&&E_2= \{ \varepsilon_{0j_A;02}
 = 0, \;\; j_A=1,\dots,7\}, \\\nonumber
&& \dots \dots   \dots \dots   \dots \dots    \\\nonumber
&& E_{N_B-1} = \{\varepsilon_{0j_A;0N_B-1}
 = 0, \;\; j_A=1,\dots,7\}.
\end{eqnarray}
Then (\ref{Cvar}) reads:
\begin{eqnarray}\label{Cvar000}
&&C=\frac{1}{2 N_B}\sqrt{\sum_{i_A=0}^{6} 
  \sum_{i_B=0}^{N_B-2} 
 \sum_{j_A=i_A+1}^{7}  
 \sum_{j_B=i_B+1}^{N_B-1}  
  \sin^2\frac{\varepsilon_{i_Aj_A;i_Bj_B}} {2}}{.}
  \end{eqnarray}
Dependence of the maximal  concurrence $C_{max}$ on the number  $N_K$ of constraint  from set (\ref{CzeroOrder3})
for  different numbers $N_B$ of  qubits in the subsystem $B$ is demonstrated in Fig.\ref{Fig:NA3}. 
Again, the maximization is performed over 1000 random choices of those  parameters   $\varepsilon_{0j_A;0j_B}$ that  remain free after imposing $N_K$  constraints from set (\ref{CzeroOrder3}), $0\le \varepsilon_{0j_A;0j_B} < 2 \pi$. 
Similar to the case $n_A=2$ in Sec.\ref{Section:Ent2qub},  the deeper jumps  down are  observed in passing from {
$N_{7 k-1}$ constraints to $N_{7k}$ constraints, $k=1,\dots, N_B-1$,}  for the same reason of introducing the linear dependence of the $k$th column of $\Phi$ (\ref{PHI2}) on the $0$th column { after adding the set of constraints $E_k$}.  {Such jumps are evident in Fig.\ref{Fig:NA3}a and are distinguishable in Fig.\ref{Fig:NA3}b for large $N_K$ (the lower parts of the curves).} 
In this case, the maximal number of operators $U_{l_Ak_B}$ controllable by the subsystem $C$ is $7 N_B$.

\begin{figure}[ht]
\centering
    \includegraphics[width=0.5\textwidth]{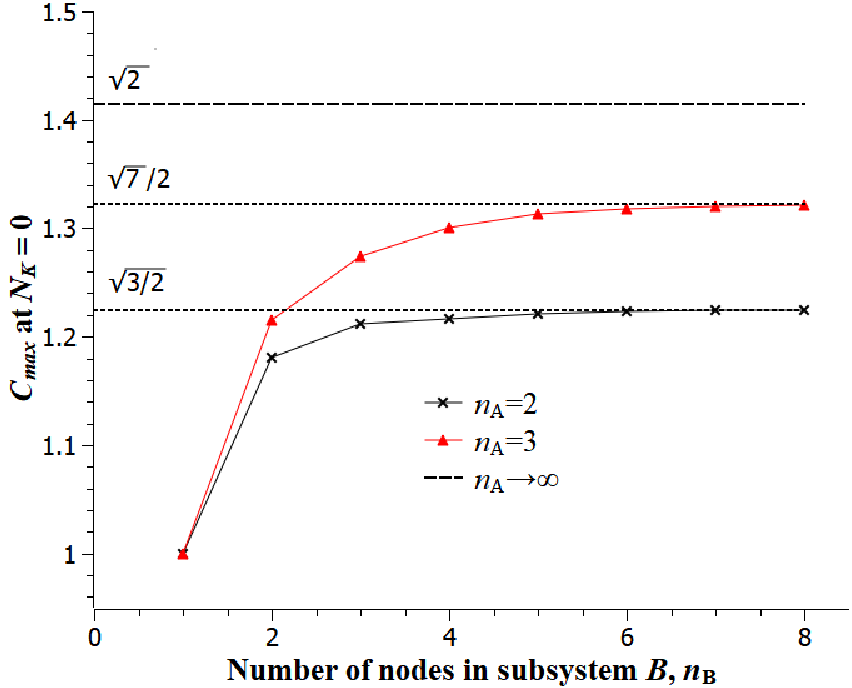}
\caption{$N_K=0$ (no constraints on the phases); the dependence of the maximal concurrence $C_{max}{(0)}$ on the number of qubits in the subsystem $B$ for two different subsystems $A$: $n_A=2,3$. The dashed lines correspond to the theoretical maximums: $C_{max}{(0)}=\;\sqrt{3/2}, \; \sqrt{7}/2,\;\sqrt{2}$ for $n_A=2,\;3,\;\infty$ respectively.}
\label{Fig:max}
\end{figure}

In Fig.\ref{Fig:max}, we show the maximal concurrences $C_{max}{(0)}$ at $N_K=0$ for $n_A=2$ and $n_A=3$ (two solid  lines)  as functions of $n_B$, {we denote them $C_{max}(n_B)$}. The horizontal dashed lines correspond to the theoretical upper limits of the concurrences associated with the maximally mixed density matrices, {$\rho_A = \frac{I_A}{N_A}$, where $I_A$ is the identity operator in the space of the subsystem $A$},  $C_{max}=\sqrt{2(1-1/N_A)}$.
For $n_A=1,\;2, \;3$, we have respectively $C_{max}=1, \;\sqrt{3/2}, \; \sqrt{7}/2$. 
 The upper dashed line corresponds to the limit $n_A\to \infty$, in this case  $C_{max} \to \sqrt{2}$. {We see that our curves $C_{max}(n_B)$ for $n_A=2$ and $3$ approach the appropriate  theoretical maximums with an  increase in $n_B$. This is due to the fact that the subsystem $B$ plays the role of  environment for the subsystem $A$ and therefore an  increase in $n_B$ leads to  an  increase in decoherence in $A$ due to the interaction with environment, so that the state of $A$ approaches the maximally mixed state. Fig.\ref{Fig:max} also demonstrates that the  concurrence $C_{max}(n_B)$ approaches the upper limit $C_{max}=\sqrt{2}$  with an increase in $n_A$.}

\section{Conclusions}
\label{Section:Conclusions}
We consider the problem of two-level control over the unitary transformations applied to a quantum system $S$ {with the purpose of creating the needed state of this system. The first-level  control center  is represented by the one-qubit subsystem $M$    (main control) that} is aimed on creating the equal-probability superposition state of the auxiliary subsystem $R$ with certain phases of probability amplitudes. More exactly, the operator $W$ imposes  certain constraints on the phases of the probability amplitudes in the superposition state of $R$ and these constraints   define the possibilities of the second-level control center $C$ {that effects the state of $R$ via the unitary $V$-operators. In other words, the number of unitary transformations $U_k$  that can be switched off by the $V$-operators  is predicted by the phases in the superposition state of $R$.}
{ We present a simplest example of two-level control scheme governing the creation of entangled state of the two-qubit system $S$.}

{It should be emphasized that we study the control  associated with the single operator (multiqubit in general)  $V_1$ acting on $n_A$ qubits of the subsystem $R$. The set of operators $V_k$, $k=1,2,\dots$, can be considered as a particular case of another single multiqubit $V$-operator acting on  all qubits subjected to action of operators $V_k$. Flexibility of such operator $V$ can be increased (i) by  replacing the single 
constraint (\ref{AlA0}) on the elements of $l_A$th row of $V$ with  $N_A-1$ constraints (\ref{AB}) and (ii) by splitting single $V$-operator into  the several independent operators with different control qubits of the subsystem $C$ together with appropriate  modifications of phase-constraint and constraints (\ref{AB}).}

It is remarkable that the above constraints on  the phases of the equal-probability state are the arguments in the sine-square functions composing expression for {the bipartite concurrence between the subsystems $A$ (to which the operator $V$ is applied) and subsystem $B$ (rest nodes)}. As the result, those  constraints  not only increase the  number of $U_k$ that can be switched off by the second-level control center $C$, but also decrease the entanglement in the bipartite system $A\cup B$.
In our model, the maximal number of operators $U_k$ controllable by multiqubit operator $V_1$   is $(N_A-1) N_B$, while  the maximal number of the constraints that can be imposed on the above phases is $(N_A-1)(N_B-1)$. 

We also present the detailed study of  the bi-partite entanglement  corresponding to one- two- and three-qubit operators  $V_1$ and demonstrate that the deep jump-down in the value of entanglement  corresponds to establishing the linear dependence of the subsequent    $k$th  column of the matrix $\Phi$ (\ref{PHI2}) on the $0$th column. On the other hand, this selects additional $N_A-1$ operators $U_{l_Ak_B}$, $l_A=0,\dots,N_A-2$, that can be switched off by the multiqubit operator $V_1$. We notice that {$l_A<N_A-1$}, otherwise $\det(V_1)=0$, as remarked in the end of Sec.\ref{Section:TLC}.

{\bf Acknowledgments.} The work was performed as a part of a state task, State Registration
No. 124013000760-0.

\section{Appendix: derivation of Eq.(\ref{Cvar}) for concurrence}
\label{Section:Appendix}
The object of our study is the pure state of an $n^{(R)}$-qubit system $R$:
\begin{eqnarray}\label{Psi0}
|\Psi\rangle = \sum_{i=0}^{2^{n^{(R)}}}  a_{i} |i\rangle, \;\;\sum_{i} |a_{i}|^2=1.
\end{eqnarray}
Let us split this system into two subsystems $A$ and $B$ including, respectively,  $n_A$ and $n_B$ nodes, and having  $N_A=2^{n_A}$ and  $N_B=2^{n_B}$ basis states:
\begin{eqnarray}\label{PsiAB}
|\Psi_{AB}\rangle = \sum_{i_A=0}^{N_A-1} \sum_{i_B=0}^{N_B-1} a_{i_Ai_B} |i_A\rangle |i_B\rangle, \;\;\sum_{i_A{,}i_B} |a_{{i_A}i_B}|^2=1.
\end{eqnarray}
Then
\begin{eqnarray}
\rho_{AB}= |\Psi_{AB}\rangle \langle \Psi_{AB}|= \sum_{i_A=0}^{N_A-1} \sum_{i_B=0}^{N_B-1} 
 \sum_{j_A=0}^{N_A-1} \sum_{j_B=0}^{N_B-1}  
  a_{i_Ai_B} a_{j_Aj_B}^*  |i_A\rangle |i_B\rangle \langle j_A| \langle j_B| .
\end{eqnarray}
Next, we calculate the partial trace over $B$:
\begin{eqnarray}\label{partTr}
\rho_A={\mbox{Tr}}_B \rho_{AB} =  \sum_{i_A=0}^{N_A-1} \sum_{i_B=0}^{N_B-1} 
 \sum_{j_A=0}^{N_A-1}  
  a_{i_Ai_B} a_{j_Ai_B}^*  |i_A\rangle \langle j_A| .
\end{eqnarray}
Then
\begin{eqnarray}
\rho_A^2=  \sum_{i_A=0}^{N_A-1} \sum_{i_B=0}^{N_B-1} 
 \sum_{j_A=0}^{N_A-1}  
 \sum_{j_B=0}^{N_B-1} 
 \sum_{k_A=0}^{N_A-1}  
  a_{i_Ai_B} a_{j_Ai_B}^* a_{j_Aj_B} a_{k_Aj_B}^* |i_A\rangle \langle k_A|{.}
\end{eqnarray}
Now, we calculate the trace of the obtained matrix
\begin{eqnarray}
{\mbox{Tr}}\rho_A^2=  \sum_{i_A=0}^{N_A-1} \sum_{i_B=0}^{N_B-1} 
 \sum_{j_A=0}^{N_A-1}  
 \sum_{j_B=0}^{N_B-1} 
  a_{i_Ai_B} a_{j_Ai_B}^* a_{j_Aj_B} a_{i_Aj_B}^* {.}
\end{eqnarray}
Let 
\begin{eqnarray}
a_{i_Ai_B} = \alpha_{i_Ai_B}e^{i \varphi_{i_Ai_B}},\;\;\; \sum_{i_A,i_B} \alpha_{i_Ai_B}^2=1,
\end{eqnarray}
where  $\alpha_{i_Ai_B}$ and $\varphi_{i_Ai_B}$ are real parameters.
We have 
\begin{eqnarray}
&&
{\mbox{Tr}}\rho_A^2=   \sum_{i_A=0}^{N_A-1} \sum_{i_B=0}^{N_B-1} 
 \sum_{j_A=0}^{N_A-1}  
 \sum_{j_B=0}^{N_B-1} \alpha_{i_Ai_B}\alpha_{i_Aj_B}\alpha_{j_Aj_B}\alpha_{j_Ai_B} \times\\\nonumber
&&
 e^{i( \varphi_{i_Ai_B}- \varphi_{i_Aj_B} +\varphi_{j_Aj_B}-\varphi_{j_Ai_B} ) }  =\\\nonumber
 &&
 \left( 
  \sum_{i_A=0}^{N_A-2} 
  \sum_{i_B=0}^{N_B-2} 
 \sum_{j_A=i_A+1}^{N_A-1}  
 \sum_{j_B=i_B+1}^{N_B-1} 
 4\alpha_{i_Ai_B}\alpha_{i_Aj_B}\alpha_{j_Aj_B}\alpha_{j_Ai_B} \times\right. \\\nonumber
&&\left. \cos( \varphi_{i_Ai_B}- \varphi_{i_Aj_B} +\varphi_{j_Aj_B}-\varphi_{j_Ai_B} )  +\right.
 \\\nonumber
 &&\left. \sum_{i_A=0}^{N_A-1}  
\sum_{j_A=0}^{N_A-1}
  \sum_{i_B=0}^{N_B-1} 
 \alpha_{i_Ai_B}\alpha_{i_Ai_B}\alpha_{j_Ai_B}\alpha_{j_Ai_B}  + \right.
 \\\nonumber
 &&\left. \sum_{i_A=0}^{N_A-1} 
   \sum_{i_B=0}^{N_B-1}  \sum_{j_B=0}^{N_B-1} 
\alpha_{i_Ai_B}\alpha_{i_Aj_B}\alpha_{i_Aj_B}\alpha_{i_Ai_B} -\right.\\\nonumber
&&\left.
 \sum_{i_A=0}^{N_A-1} 
   \sum_{i_B=0}^{N_B-1}  
\alpha_{i_Ai_B}\alpha_{i_Ai_B}\alpha_{i_Ai_B}\alpha_{i_Ai_B}
  \right.
 \Big)  .
\end{eqnarray}
If all $\alpha_{i_Ai_B}=\frac{1}{\sqrt{N_AN_B}}$ (equal probability state), then this formula becomes simpler 
\begin{eqnarray}
&&
{\mbox{Tr}}\rho_A^2= \frac{1}{N_A^2N_B^2}  \sum_{i_A=0}^{N_A-1} \sum_{i_B=0}^{N_B-1} 
 \sum_{j_A=0}^{N_A-1}  
 \sum_{j_B=0}^{N_B-1} 
 e^{i( \varphi_{i_Ai_B}- \varphi_{i_Aj_B} +\varphi_{j_Aj_B}-\varphi_{j_Ai_B} ) }  =\\\nonumber
 &&
  \frac{1}{N_A^2N_B^2} \left( 
  \sum_{i_A=0}^{N_A-2} 
  \sum_{i_B=0}^{N_B-2} 
 \sum_{j_A=i_A+1}^{N_A-1}  
 \sum_{j_B=i_B+1}^{N_B-1} 
 4 \cos( \varphi_{i_Ai_B}- \varphi_{i_Aj_B} +\varphi_{j_Aj_B}-\varphi_{j_Ai_B} )  +\right.
 \\\nonumber
 &&\left.
 N_AN_B (N_A+N_B-1)\right. \Big)=\\\nonumber
 &&
 1 - \frac{8}{N_A^2N_B^2} \left(
 \sum_{i_A=0}^{N_A-2} 
  \sum_{i_B=0}^{N_B-2} 
 \sum_{j_A=i_A+1}^{N_A-1}  
 \sum_{j_B=i_B+1}^{N_B-1}  
  \sin^2\frac{\varphi_{i_Ai_B}- \varphi_{i_Aj_B} +\varphi_{j_Aj_B}-\varphi_{j_Ai_B}} {2}\right)  .
\end{eqnarray}
As the result,  formula (\ref{C}) yields expression (\ref{Cvar}) for the bi-partite concurrence in the equal-probability state.


\begin{thebibliography}{99}

\bibitem{KRKS}
N. Khaneja, T. Reiss, C. Kehlet, T. Schulte-Herbr\"uggen, and S.J. Glaser, 
Optimal control of coupled spin dynamics: design of NMR pulse sequences
by gradient ascent algorithms,  J. Magn. Reson. {\bf 172}(2), 296 (2005)

\bibitem{PP_2023}
 V.N. Petruhanov and A.N. Pechen, GRAPE optimization for open quantum
systems with time-dependent decoherence rates driven by coherent
and incoherent controls, J. Phys. A: Math. Theor. {\bf 56}(30), 305303 (2023)

\bibitem{L}
 D. Lucarelli, Quantum optimal control via gradient ascent in function space
and the time-bandwidth quantum speed limit,  Phys. Rev. A {\bf 97}(6), 062346 (2018)

\bibitem{GV}
D.L. Goodwin, and M.S. Vinding,  Accelerated Newton-Raphson GRAPE
methods for optimal control, Phys. Rev. Res. {\bf 5}(1), L012042 (2023)

\bibitem{FSGK}
P. De Fouquieres, S.G. Schirmer, S.J. Glaser, and I. Kuprov, Second order
gradient ascent pulse engineering, J. Magn. Reson. {\bf 212}(2), 412 (2011)

\bibitem{KF}
V.F. Krotov,  and N.N. Fel'dman,  Iterative Method to Solve the Problems of Optimal Control, Izv. Akad. Nauk SSSR, Tekh. Kibern. no. 2, 160 (1983)

\bibitem{K}
Krotov, V.F., Global Methods in Optimal Control Theory, New York: Marcel Dekker (1996)

\bibitem{MP_2019}
O.V. Morzhin and A.N. Pechen, Krotov method for optimal control of closed quantum
systems, Russian Math. Surveys {\bf 74}(5), 851-908 (2019)

\bibitem{FFLC}
M. Fernandes, F.F. Fanchini, E. de Lima, and L.K. Castelano, Effectiveness of the
Krotov method in finding controls  for open quantum systems, J.Phys.A:Mth.Theor. {\bf 56}(49) (2024)

\bibitem{BM}
O.V. Baturina and O.V. Morzhin, Optimal control of the spin system on a basis of
the global improvement method, Autom. Remote Control {\bf 72}(6), 1213 (2011)

\bibitem{SJL}
S.G. Schirmer, E.A. Jonckheere, and F.C. Langbein, Design of feedback control
laws for information transfer in spintronics networks, IEEE Trans. Autom. Control
{\bf 63}(8), 2523 (2018)

\bibitem{MW}
S. Mancini and H.M. Wiseman,
Optimal control of entanglement via quantum
feedback, Phys. Rev. A {\bf 75}(1), 012330 (2007)

\bibitem{BWPRWL}
 J. Biamonte, P. Wittek, N. Pancotti, P. Rebentrost, N. Wiebe, and
S. Lloyd, Quantum machine learning, Nature {\bf 549}, 195 (2017)

\bibitem{NBSN}
 M.Y. Niu, S. Boixo, V. Smelyanskiy, and H. Neven, Universal quantum control
through deep reinforcement learning, npj Quantum Information {\bf 5}, 33 (2019)

\bibitem{GSB}
L. Giannelli, P. Sgroi, J. Brown, G. Paraoanu, M. Paternostro, E. Paladino, G. Falci, A tutorial on optimal control and reinforcement
learning methods for quantum technologies, Phys. Lett. A {\bf 434}, 128054
(2022)

{
\bibitem{CZCW}
Sh.-J. Cao, L.-N. Zheng, L.-Yo. Cheng
 and H.-F. Wang,
Controllable entangled-state transmission in a non-Hermitian trimer Su-Schrieffer-Heeger chain,
Phys.Rev.A {\bf 110}, 062409 (2024)

\bibitem{NJS}
T. Neema, S. Jha, T.Sahai,
Non-Markovian quantum control via model maximum likelihood estimation and reinforcement learning, 
	arXiv:2402.05084 (2024)

\bibitem{CBMF}
M.F. Cavalcante, M. V. S. Bonanca, E. Miranda, and S. Deffner,
Nano-welding of quantum spin-1/2 chains at minimal dissipation,
Phys.Rev.B {\bf 110}, 064304 (2024)

\bibitem{CLY}
P.-Ju Chen, D.-W. Luo, and T. Yu,
Optimal entanglement generation in optomechanical systems via Krotov control of covariance matrix dynamics,
Phys.Rev. Research, accepted  (2025)

\bibitem{LFHJ}
C.-Y. Liu, C. G. Feyisa, M.S. Hasan, and H.H. Jen,
High-fidelity multipartite entanglement creation in non-Hermitian qubits, 
arXiv:2412.01133v1 [quant-ph] (2024)

\bibitem{M}
M. Mafu,
Advances in artificial intelligence and machine learning for
quantum communication applications, IET Quantum Communication
{\bf 5}(3), 202 (2024)
}


\bibitem{PBGWK}
Peters N.A., Barreiro J.T., Goggin M.E., Wei T.-C., and Kwiat P.G. Remote State
Preparation: Arbitrary Remote Control of Photon Polarization, Phys.Rev.Lett. {\bf 94},
(2005) 150502

\bibitem{PBGWK2}
Peters N.A., Barreiro J.T., Goggin M.E., Wei T.-C., and Kwiat P.G. Remote State
Preparation: Arbitrary remote control of photon polarizations for quantum communication,
Quantum Communications and Quantum Imaging III, ed. R.E.Meyers, Ya.Shih. Proc. of
SPIE  {\bf 5893} SPIE. Bellingham. WA. (2005)

\bibitem{DLMRKBPVZBW}
Dakic B. , Lipp Ya.O., Ma X., Ringbauer M., Kropatschek S., Barz S., Paterek T., Vedral
V., Zeilinger A., Brukner C., and Walther P. Quantum discord as resource for remote state
preparation, Nat. Phys. {\bf 8}, (2012) 666

\bibitem{PSB}
Pouyandeh S., Shahbazi F., Bayat A. Measurement-induced dynamics for spin-chain quantum
communication and its application for optical lattices, Phys.Rev.A. {\bf 90}, 
(2014) 012337

\bibitem{LH}
Liu L.L., and Hwang T. Controlled remote state preparation protocols via AKLT states,
Quant.Inf.Process. {\bf 13} (2014) 1639

\bibitem{Z_2014}
Zenchuk A.I. Remote  creation of a one-qubit mixed state through a short homogeneous spin-1/2 chain,
Phys.Rev.A. {\bf 90} (2014) 052302

\bibitem{BZ_2015}
Bochkin G.A., and Zenchuk A.I. Remote one-qubit-state control using the pure initial state of a two-qubit sender:
Selective-region and eigenvalue creation,  Phys.Rev.A. {\bf 91}, (2015) 062326

\bibitem{SW}
M. Sawerwain, J. Wi\'sniewska,  Quantum Switch Realization by the Quantum Lyapunov Control. In: Gaj, P., Sawicki, M., Kwiecie\'n, A. (eds) Computer Networks. CN 2019. Communications in Computer and Information Science, vol 1039. Springer, Cham. (2019)

\bibitem{Licina}
 M. Licina
High-dimensional quantum state manipulation and tracking,
 International Journal of Quantum Information
{\bf 18}(7)  2050045  (2020) 

\bibitem{JWQCCLXSZM}
X. Jiang, K. Wang, K. Qian,
 Zh. Chen, Zh. Chen, L. Lu, L. Xia, F. Song, Sh. Zhu, X. Ma,
Towards the standardization of quantum state verification using optimal strategies,  npj Quantum Inf. {\bf 6}, 90 (2020)

\bibitem{FZ_2017}
E.B.Fel'dman and A.I.Zenchuk Coherence Evolution and Transfer Supplemented
by Sender's Initial-State Restoring,  JETP {\bf 125}(6), (2017) 1042

\bibitem{FPZ_2021}
E.B. Fel'dman, A.N. Pechen, A.I. Zenchuk, Complete structural restoring of transferred multi-qubit quantum state, 
Phys.Lett. A {\bf 413} (2021) 127605

\bibitem{BFLPZ_2022}
 G.A.Bochkin, E.B.Fel'dman, I.D.Lazarev, A.N.Pechen,
A.I.Zenchuk, Transfer of zero-order coherence matrix along spin-1/2
chain, Quant.Inf.Proc. {\bf 21}, 261
(2022)

\bibitem{FPZ_2024}
 E.B.Fel'dman, A.N. Pechen and A.I.Zenchuk, Optimal remote restoring of quantum states in communication lines via  
local magnetic field, Phys. Scr. {\bf  99}, 025112 (2024) 

 \bibitem{NC}
Nielsen M.A., and Chuang I.L. Quantum Computation and Quantum Information.
Cambridge:Cambridge Univ.Press. (2010) 

\bibitem{BBPS}
H. Bennett, H. J. Bernstein, S. Popescu, and B. Schumacher, Phys. Rev. A {\bf 53}, 2046 (1996)


\bibitem{HW}
S.Hill, and W.K.Wootters, Entanglement of a pair of quantum bits, Phys.Rev.Lett. {\bf 78}, 5022 (1997)

\bibitem{Wootters}
 W.K.Wootters, Entanglement of formation of an arbitrary state of two qubits, Phys.Rev.Lett. {\bf 80}, 2245 (1998) 
 
\bibitem{RBCHM}
P. Rungta, V. Bu\v{z}ek, C.M. Caves, M. Hillery, and G. J. Milburn,
Universal state inversion and concurrence in arbitrary dimensions, Phys.Rev.A {\bf 64}, 042315 (2001)


\end{thebibliography}
\end{document}